\documentclass[12pt]{article}
\usepackage{graphicx}
\begin{document}

\begin{center}
{\LARGE \bf Feynman's Branes and Feynman's Oscillators}

\vspace{3ex}

Y. S. Kim\footnote{electronic address: yskim@physics.umd.edu}\\
Department of Physics, University of Maryland,\\
College Park, Maryland 20742, U.S.A.\\

\vspace{3ex}

Marilyn E. Noz \footnote{electronic address: noz@nucmed.med.nyu.edu}\\
Department of Radiology, New York University,\\ New York,
New York 10016, U.S.A.

\end{center}

\vspace{3ex}

\begin{abstract}

Based on Feynman's lifetime efforts on quantum mechanics and relativity,
it is concluded that the basic difference between field theory
and string theory is that field theory is based on running waves while
string theory should deal with standing waves in the Lorentz-covariant
regime.

At the 1970 spring meeting of the American Physical Society held in
Washington, DC, R. P. Feynman stunned the audience by proposing
harmonic oscillators for relativistic bound states, instead of
Feynman diagrams.  His talk was later published in the paper of
Feynman, Kislinger, and Ravndal [Phys. Rev. D, Vol. 3, 2706 (1971)].
These authors noted that the hadron mass spectra can be predicted
by the degeneracy of the three-dimensional harmonic oscillators.
In so doing, they started with the Lorentz-invariant differential
equation for the harmonic oscillator, and obtained Lorentz invariant
solutions.  However, their solutions are not normalizable in the
time-separation variable and cannot carry probability interpretation.
It is pointed out that there are solutions normalizable in the
time-separation variable within the framework of Wigner's little-group
representation of the Poincar\'e group.  These solutions are not
invariant but covariant under Lorentz transformations.  These solutions
give a covariant bound-state model which gives the quark model and
the parton model as two different limiting cases, in the low- and
high-speed limits respectively.

\end{abstract}

\newpage

\section{Inroduction}
Isaac Newton discovered the inverse-square law between two point particles, such
as the sun and earth. It took him twenty years to formulate the gravity law
for non-point particles such as the sun and earth.  He had to invent a new
mathematics to solve the problem of extended particles.  We all know what
mathematics Newton had to invent.

Ninety-nine years ago, Einstein formulated relativistic dynamics for point
particles.  Since then, particles became entities in quantum world.  Yet,
Einstein's energy-momentum relation prevails for all those relativistic
particles.  The question then is the internal variables.  In quantum world,
particles have intrinsic spins.  Are those spins consistent with
Einstein's Lorentz covariance?   This question was raised by Wigner in his
1939 paper on representations of the Poincar\'e group~\cite{wig39}.

\begin{table}[thb]
\caption{Massive and massless particles in one package.  Wigner's
little group unifies the internal space-time symmetries for massive and
massless particles.  It is a great challenge for us to find
another unification: the unification of the quark and parton pictures in
high-energy physics.}\label{table11}
\vspace{3mm}                                                   \begin{center}
\begin{tabular}{lccc}
\hline
{}&{}&{}&{}\\
{} & Massive, Slow \hspace{6mm} & COVARIANCE \hspace{6mm}&
Massless, Fast \\[4mm]\hline
{}&{}&{}&{}\\
Energy- & {}  & Einstein's & {} \\
Momentum & $E = p^{2}/2m$ & $ E = [p^{2} + m^{2}]^{1/2}$ & $E = p$
\\[4mm]\hline
{}&{}&{}&{}\\
Internal & $S_{3}$ & {}  &  $S_{3}$ \\[-1mm]
Space-time &{} & Wigner's  & {} \\ [-1mm]
Symmetry & $S_{1}, S_{2}$ & Little Group & Gauge Trans. \\[4mm]\hline
{}&{}&{}&{}\\
Relativistic & {} & One  &  {} \\[-1mm]
Extended & Quark Model & Covariant  & Parton Model\\ [-1mm]
Particles & {} & Theory &{} {} \\[4mm]\hline

\end{tabular}

\end{center}
\end{table}

Table~\ref{table11} summarizes the covariant picture of the present
particle world.  The second row of this table indicates that the spin
symmetry of slow particles and the helicity-gauge symmetry of massless
particles are two limiting cases of one covariant entity called Wigner's
little group.  This issue has been extensively discussed in the
literature~\cite{kiwi90jm}.

Let us then concentrate on the third row of Table~\ref{table11}.  After
Einstein formulated his special relativity, a pressing problem was to see
whether his relativistic dynamics can be extended to rigid bodies as
in the case of Newton's sun and earth and their rotations.  As far as
we know, there are no statisfactory solutions to this problem.  However,
according to quantum mechanics, these extended objects are wave packets
or standing waves.  It might to easier to deal with waves in
Einstein's relativistic world.

As we all know, it is a trivial matter to write down plane waves in a
covariant manner. The expression is Lorentz-invariant.  Indeed, quantum
field theory is possible because the plane waves take an invariant form.
Plane waves are running waves.

As for localized objects, like the sun or earth, we have to consider
standing waves.  Standing waves are superposition of running waves in
opposite directions.  Do those superpositions remain invariant under
Lorentz boosts?  This is the question we wish to exploit from the
1971 paper of Feynman, Kislinger and Ravandal~\cite{fkr71}.

In discussing standing waves, it is common to start with a hard-wall
potential, but mathematically it is more comfortable to use harmonic
oscillators.  Indeed, in their paper of 1971~\cite{fkr71}, Feynman
{\it et al.} start with a Lorentz-invariant harmonic oscillator equation.
This equation has many different solutions satisfying different boundary
conditions.  The solution they use is not normalizable in time-separation
variable, and cannot be given any physical interpretation.

On the other hand, it is possible to fix up their mathematics.  Their
Lorentz-invariant differential equation has a normalizable solution
which can form a representation space for Wigner's little group for
massive particles~\cite{knp86}.  We shall discuss this solution in
more detail in this report.

Then, there is another question.  Quantum field theory gives a calculational
tool called the S-matrix.  In the early 1960s, there was a movement in
the physics community to start with the S-matrix, instead of complicated
field theory, in spite of its field theoretic origin.  This proposal was
based on analytic properties of the S-matrix, and on the premise that the
physics can be formulated in terms of singularities in the complex plane.

While the S-matrix deals primarily with scattering states, bound states
can be found from the poles in the negative-energy region.  This is
indeed the case for nonrelativistic potential scattering.  Therefore
there was a movement to understand bound-state problems using
analytic properties of the S-matrix.

In order to see what happened in this approach, we shall discuss in
Sec.~\ref{dff} a concrete example of the neutron-proton mass difference
which once showed a promise but which did not work out.  This case is
known as the Dashen-Frautschi fiasco in the physics community.  We shall
show that this fiasco was caused by the confusion about running and
standing waves in quantum mechanics. In Sec~\ref{waves}, we point out
there are running waves and standing waves in quantum mechanics.  While
it is trivial to Lorentz-boost running waves, it requires covariance of
boundary conditions to understand fully standing waves.
In Sec.~\ref{covham}, we construct the covariant harmonic oscillator
wave functions.  These wave functions can be Lorentz-boosted, but
they depend on the time-separation variable.  It is shown in Sec.~\ref{par}.
the quark and parton models are two different manifestation of the
same covariant entity.  The most controversial aspect of Feynman's
parton picture is that the partons interact incoherently with external
signals.

\section{Dashen-Frautschi Fiasco}\label{dff}

On April 29, at the 1965 spring meeting of the American Physical
Society in Washington, Freeman J. Dyson of the Institute of Advanced
Study (Princeton) presented an invited talk entitled "Old and New
Fashions in Field Theory," and the content of his talk
was published in the June issue of the Physic Today, on page 21-24
(1965).  This paper contains the following paragraph.

\begin{quote}
The first of these two achievements is the explanation
of the mass difference between neutron
and proton by Roger Dashen, working at the time as a graduate
student	under the supervision of Steve Frautschi.
The neutron-proton mass difference has for thirty years	been believed
to be electromagnetic in origin, and it offers a splendid
experimental test of any theory	which tries to cover the borderline
between electromagnetic and strong	interactions.	However,
no convincing theory of the mass-difference had appeared before
1964.  In this connection I exclude as unconvincing all	theories,
like the early	theory	of Feynman and Speisman, which use one
arbitrary cut-off parameter to	fit one	experimental number.
Dashen	for the first time made an honest calculation without arbitrary
parameters and got the right answer.  His method is a
beautiful marriage between old-fashioned electrodynamics and modern
bootstrap techniques.  He writes down the equations expressing
the fact that the neutron can be considered to be a bound state of a
proton with a negative pi meson, and the proton	a bound state of
a neutron with a positive pi meson, according to the bootstrap method.
Then into these equations he puts electromagnetic perturbations, the
interaction of a photon with both nucleon and pi meson, according to
the Feynman rules.  The calculation of the resulting mass difference is
neither long nor hard to understand, and in my opinion, it will become
a classic in the history of physics.
\end{quote}

Dyson was talking about the paper by R. F. Dashen and S. C. Frautschi
published in the Physical Review~\cite{df64}.  They use
the S-matrix formalism for bound states. In their paper, Dashen and
Frautschi use the S-matrix method to calculate a perturbed energy level.
Of course, they use approximations because they are dealing with strong
interactions.  There are however ``good'' approximations and ``bad''
approximations.

\begin{figure}[thb]
\centerline{\includegraphics[scale=0.6]{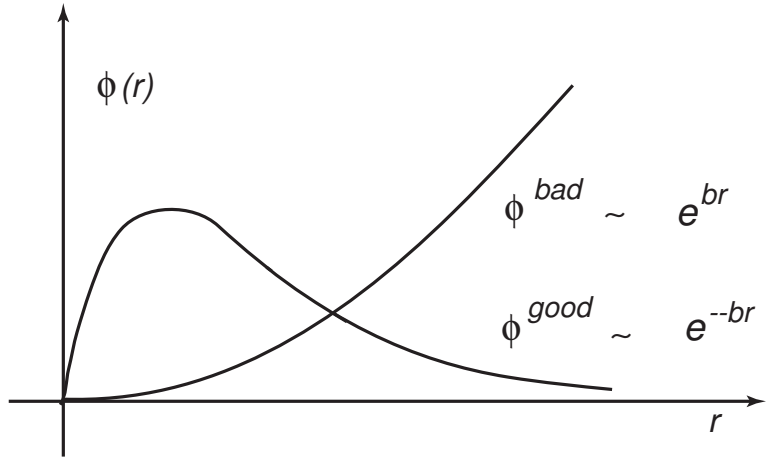}}
\vspace{5mm}
\caption{Good and bad wave functions from the S-matrix theory.  Bound-state
wave functions satisfy the localization condition and are good wave functions.
Analytic continuations of plane waves do not satisfy the localization
boundary condition, and become bad wave functions at the bound-state
energy.}\label{goodbad}
\end{figure}

If we translate what they did into the language of the Schr\"odinger
picture, they are using the following approximation for for the
bound-state energy shift~\cite{kim66}
\begin{equation}\label{shift}
\delta E = \left(\phi^{good}, \delta V \phi^{bad} \right) ,
\end{equation}
where
\begin{eqnarray}\label{wfs}
&{}& \phi^{good} \sim e^{-br} , \nonumber \\[2ex]
&{}& \phi^{bad} \sim e^{br} ,
\end{eqnarray}
as illustrated in Fig.~\ref{goodbad}.

The Schr\"odinger equation is a second-order differential equation
with two solutions.  If the energy positive, there are two running-wave
solutions.  For negative energies, the two solutions take ``good''
and ``bad" forms as indicated in Eq.(\ref{wfs}).  The good wave function
is normalizable and carries probability interpretation.  The bad wave
function is not normalizable, and cannot be given any physical
interpretation.  If we demand that this bad wave function disappear,
energy-levels become discrete.  This is how the bound-state energy
levels are quantized.

In the S-matrix formalism, the bound-states appear as poles in the complex
energy plane.  Those bound-state poles correspond to``good'' localized
wave functions in the Schr\"odinger picture.  At all other places, there
are unlocalized ``bad'' wave functions.  Dashen and Frautschi overlooked
this point when they used approximations in the S-matrix theory, and
ended up with the ``bad'' formula given in Eq.(\ref{shift}).

\section{Running Waves and Standing Waves}\label{waves}
The Dashen-Frautschi fiasco teaches us an important lesson.  There are
running waves and standing waves in quantum mechanics.  Even though
the standing wave is a superposition of running waves, it requires
an additional care of boundary conditions.  We do not know how to
deal with this problem in the S-matrix formalism.

If not impossible, it is very difficult to formulate Lorentz boosts
for rigid bodies.  On the other hand, it seems to be feasible to
boost waves. Indeed, quantum mechanics allows us to look at extended
object as wave packets or standing waves.  Thus, we are interested
in boosting waves.  We should note here also that there are standing
and running waves.

Plane waves are running waves.  It is trivial to Lorentz-boost the
plane wave of the form
\begin{equation}
\exp{\left\{i\left(\vec{p}\cdot \vec{x} - p_{0} t \right)\right\}},
\end{equation}
because the exponent is invariant under Lorentz transformations.
However, what would happen when different waves are superposed?
Would the spectral function be covariant or invariant?
What would happen for standing waves which consist of superposition
of waves moving in opposite directions?

While quantum field theory based on Feynman diagrams starts with
running waves, quantum mechanics within a localized space-time region
deals with standing waves.  If string theory is set to solve the
problem inside particles, the physics of string theory is necessarily
the quantum mechanics of standing waves.

In an attempt to obtain the answers to these questions, we can start
with some examples.  As usual in quantum mechanics, the first example
for standing waves should be a set of harmonic oscillator wave
functions.  With this point in mind, let us see what Feynman did for
harmonic oscillators in the relativistic regime.

\section{Can harmonic oscillators be made covariant?}\label{covham}

Quantum field theory has been quite successful in terms of
perturbation techniques in quantum electrodynamics.  However, this
formalism is based on the S matrix for scattering problems and useful
only for physical processes where a set of free particles becomes
another set of free particles after interaction.  Quantum field theory
does not address the question of localized probability distributions
and their covariance under Lorentz transformations.
The Schr\"odinger quantum mechanics of the hydrogen atom deals with
localized probability distribution.  Indeed, the localization condition
leads to the discrete energy spectrum.  Here, the uncertainty relation
is stated in terms of the spatial separation between the proton and
the electron.  If we believe in Lorentz covariance, there must also
be a time-separation between the two constituent particles.

\begin{figure}[thb]
\centerline{\includegraphics[scale=1.0]{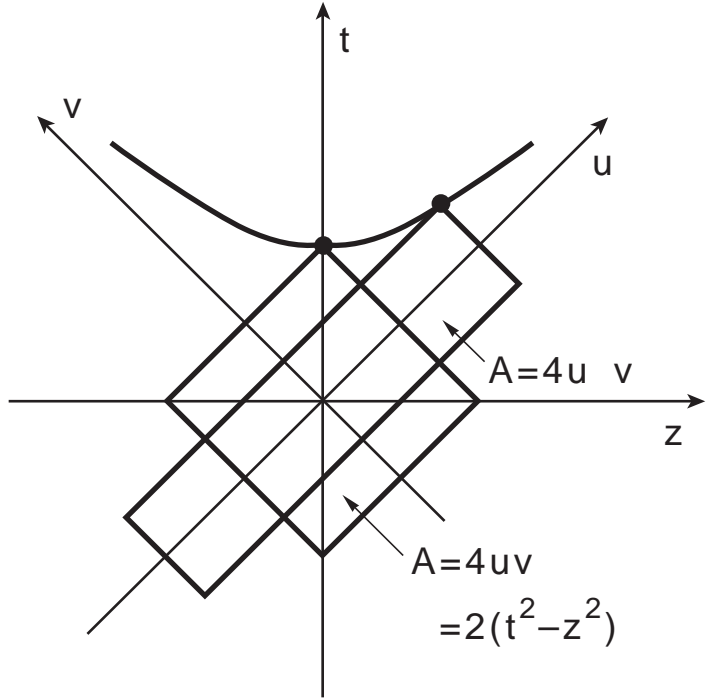}}
\vspace{5mm}
\caption{Lorentz boost in the light-cone coordinate
system.}\label{licone}
\end{figure}

Before 1964~\cite{gell64}, the hydrogen atom was used for
illustrating bound states.  These days, we use hadrons which are
bound states of quarks.  Let us use the simplest hadron consisting of
two quarks bound together with an attractive force, and consider their
space-time positions $x_{a}$ and $x_{b}$, and use the variables
\begin{equation}
X = (x_{a} + x_{b})/2 , \qquad x = (x_{a} - x_{b})/2\sqrt{2} .
\end{equation}
The four-vector $X$ specifies where the hadron is located in space and
time, while the variable $x$ measures the space-time separation
between the quarks.  According to Einstein, this space-time separation
contains a time-like component which actively participates as can be
seen from
\begin{equation}\label{boostm}
\pmatrix{z' \cr t'} = \pmatrix{\cosh \eta & \sinh \eta \cr
\sinh \eta & \cosh \eta } \pmatrix{z \cr t} ,
\end{equation}
when the hadron is boosted along the $z$ direction.
In terms of the light-cone variables defined as~\cite{dir49}
\begin{equation}
u = (z + t)/\sqrt{2} , \qquad v = (z - t)/\sqrt{2} ,
\end{equation}
the boost transformation of Eq.(\ref{boostm}) takes the form
\begin{equation}\label{lorensq}
u' = e^{\eta } u , \qquad v' = e^{-\eta } v .
\end{equation}
The $u$ variable becomes expanded while the $v$ variable becomes
contracted, as is illustrated in Fig.~\ref{licone}.

Does this time-separation variable exist when the hadron is at rest?
Yes, according to Einstein.  In the present form of quantum mechanics,
we pretend not to know anything about this variable.  Indeed, this
variable belongs to Feynman's rest of the universe.  In this report,
we shall see the role of this time-separation variable in the
decoherence mechanism.

\begin{figure}[thb]
\centerline{\includegraphics[scale=1.0]{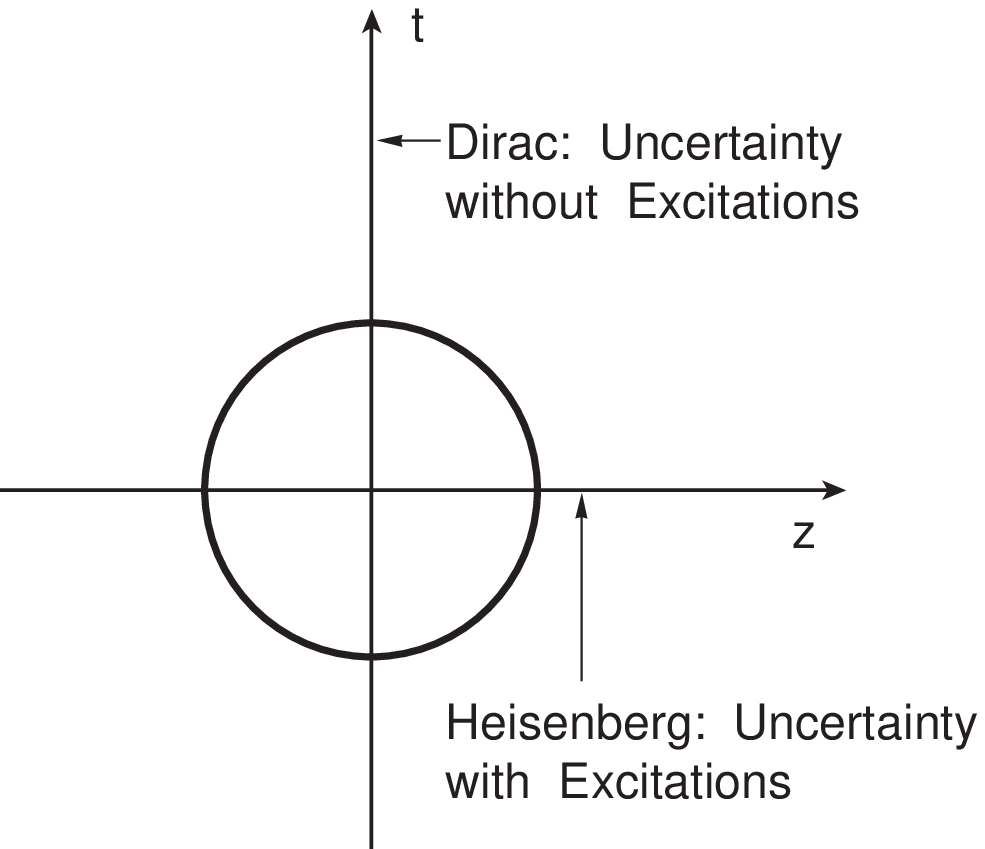}}
\vspace{5mm}
\caption{Space-time picture of quantum mechanics.  There
are quantum excitations along the space-like longitudinal direction, but
there are no excitations along the time-like direction.  The time-energy
relation is a c-number uncertainty relation.}\label{quantum}
\end{figure}

Also in the present form of quantum mechanics, there is an uncertainty
relation between the time and energy variables.  However, there are
no known time-like excitations.  Unlike Heisenberg's
uncertainty relation applicable to position and momentum, the time and
energy separation variables are c-numbers, and we are not allowed to
write down the commutation relation between them.  Indeed, the
time-energy uncertainty relation is a c-number uncertainty
relation~\cite{dir27}, as is illustrated in Fig.~\ref{quantum}

How does this space-time asymmetry fit into the world of
covariance~\cite{kn73}.  This question was studied in depth by the
present authors in the past.  The answer is that Wigner's $O(3)$-like
little group is not a Lorentz-invariant symmetry, but is a covariant
symmetry~\cite{wig39}.  It has been shown that the time-energy
uncertainty applicable to the time-separation variable fits perfectly
into the $O(3)$-like symmetry of massive relativistic
particles~\cite{knp86}.

The c-number time-energy uncertainty relation allows us to write down
a time distribution function without excitations~\cite{knp86}.
If we use Gaussian forms for both space and time distributions, we
can start with the expression
\begin{equation}\label{ground}
\left({1 \over \pi} \right)^{1/2}
\exp{\left\{-{1 \over 2}\left(z^{2} + t^{2}\right)\right\}}
\end{equation}
for the ground-state wave function.  What do Feynman {\it et al.}
say about this oscillator wave function?

\begin{figure}[thb]
\centerline{\includegraphics[scale=0.6]{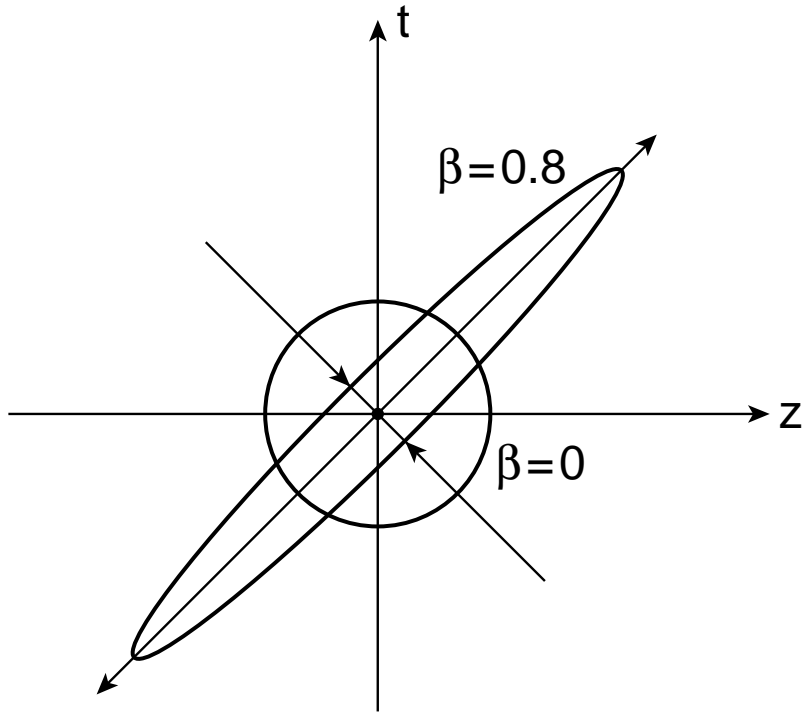}}
\caption{Effect of the Lorentz boost on the space-time
wave function.  The circular space-time distribution at the rest frame
becomes Lorentz-squeezed to become an elliptic
distribution.}\label{ellipse}
\end{figure}

In their classic 1971 paper~\cite{fkr71}, Feynman {\it et al.} start
with the following Lorentz-invariant differential equation.
\begin{equation}\label{osceq}
{1\over 2} \left\{x^{2}_{\mu} -
{\partial^{2} \over \partial x_{\mu }^{2}}
\right\} \psi(x) = \lambda \psi(x) .
\end{equation}
This partial differential equation has many different solutions
depending on the choice of separable variables and boundary conditions.
Feynman {\it et al.} insist on Lorentz-invariant solutions which are
not normalizable.  On the other hand, if we insist on normalization,
the ground-state wave function takes the form of Eq.(\ref{ground}).
It is then possible to construct a representation of the
Poincar\'e group from the solutions of the above differential
equation~\cite{knp86}.  If the system is boosted, the wave function
becomes
\begin{equation}\label{eta}
\psi_{\eta }(z,t) = \left({1 \over \pi }\right)^{1/2}
\exp\left\{-{1\over 2}\left(e^{-2\eta }u^{2} +
e^{2\eta}v^{2}\right)\right\} .
\end{equation}
This wave function becomes Eq.(\ref{ground}) if $\eta$ becomes zero.
The transition from Eq.(\ref{ground}) to Eq.(\ref{eta}) is a
squeeze transformation.  The wave function of Eq.(\ref{ground}) is
distributed within a circular region in the $u v$ plane, and thus
in the $z t$ plane.  On the other hand, the wave function of
Eq.(\ref{eta}) is distributed in an elliptic region with the light-cone
axes as the major and minor axes respectively.  If $\eta$ becomes very
large, the wave function becomes concentrated along one of the
light-cone axes.  Indeed, the form given in Eq.(\ref{eta}) is a
Lorentz-squeezed wave  function.  This squeeze mechanism is
illustrated in Fig.~\ref{ellipse}.

There are many different solutions of the Lorentz invariant differential
equation of Eq.(\ref{osceq}).  The solution given in Eq.(\ref{eta})
is not Lorentz invariant but is covariant.  It is normalizable
in the $t$ variable, as well as in the space-separation variable $z$.
How can we extract probability interpretation from this covariant
wave function?


\section{Feynman's Parton Picture}\label{par}

It is a widely accepted view that hadrons are quantum bound states
of quarks having localized probability distribution.  As in all
bound-state cases, this localization condition is responsible for
the existence of discrete mass spectra.  The most convincing evidence
for this bound-state picture is the hadronic mass spectra which are
observed in high-energy laboratories~\cite{fkr71,knp86}.

In 1969, Feynman observed that a fast-moving hadron can be regarded
as a collection of many ``partons'' whose properties appear to be
quite different from those of the quarks~\cite{fey69}.  For example,
the number of quarks inside a static proton is three, while the number
of partons in a rapidly moving proton appears to be infinite.  The
question then is how the proton looking like a bound state of quarks
to one observer can appear different to an observer in a different
Lorentz frame?  Feynman made the following systematic observations.

\begin{itemize}

\item[a.]  The picture is valid only for hadrons moving with
  velocity close to that of light.

\item[b.]  The interaction time between the quarks becomes dilated,
   and partons behave as free independent particles.

\item[c.]  The momentum distribution of partons becomes widespread as
   the hadron moves fast.

\item[d.]  The number of partons seems to be infinite or much larger
    than that of quarks.

\end{itemize}

\noindent Because the hadron is believed to be a bound state of two
or three quarks, each of the above phenomena appears as a paradox,
particularly b) and c) together.

In order to resolve this paradox, let us write down the
momentum-energy wave function corresponding to Eq.(\ref{eta}).
If the quarks have the four-momenta $p_{a}$ and $p_{b}$, we can
construct two independent four-momentum variables~\cite{fkr71}
\begin{equation}
P = p_{a} + p_{b} , \qquad q = \sqrt{2}(p_{a} - p_{b}) ,
\end{equation}
where $P$ is the total four-momentum and is thus the hadronic
four-momentum.

$q$ measures the four-momentum separation between
the quarks.  Their light-cone variables are
\begin{equation}\label{conju}
q_{u} = (q_{0} - q_{z})/\sqrt{2} ,  \qquad
q_{v} = (q_{0} + q_{z})/\sqrt{2} .
\end{equation}
The resulting momentum-energy wave function is
\begin{equation}\label{phi}
\phi_{\eta }(q_{z},q_{0}) = \left({1 \over \pi }\right)^{1/2}
\exp\left\{-{1\over 2}\left(e^{-2\eta}q_{u}^{2} +
e^{2\eta}q_{v}^{2}\right)\right\} .
\end{equation}
Because we are using here the harmonic oscillator, the mathematical
form of the above momentum-energy wave function is identical to that
of the space-time wave function.  The Lorentz squeeze properties of
these wave functions are also the same.  This aspect of the squeeze
has been exhaustively discussed in the
literature~\cite{knp86,kn77par,kim89}.

When the hadron is at rest with $\eta = 0$, both wave functions
behave like those for the static bound state of quarks.  As $\eta$
increases, the wave functions become continuously squeezed until
they become concentrated along their respective positive
light-cone axes.  Let us look at the z-axis projection of the
space-time wave function.  Indeed, the width of the quark distribution
increases as the hadronic speed approaches that of the speed of
light.  The position of each quark appears widespread to the observer
in the laboratory frame, and the quarks appear like free particles.

\begin{figure}
\centerline{\includegraphics[scale=0.5]{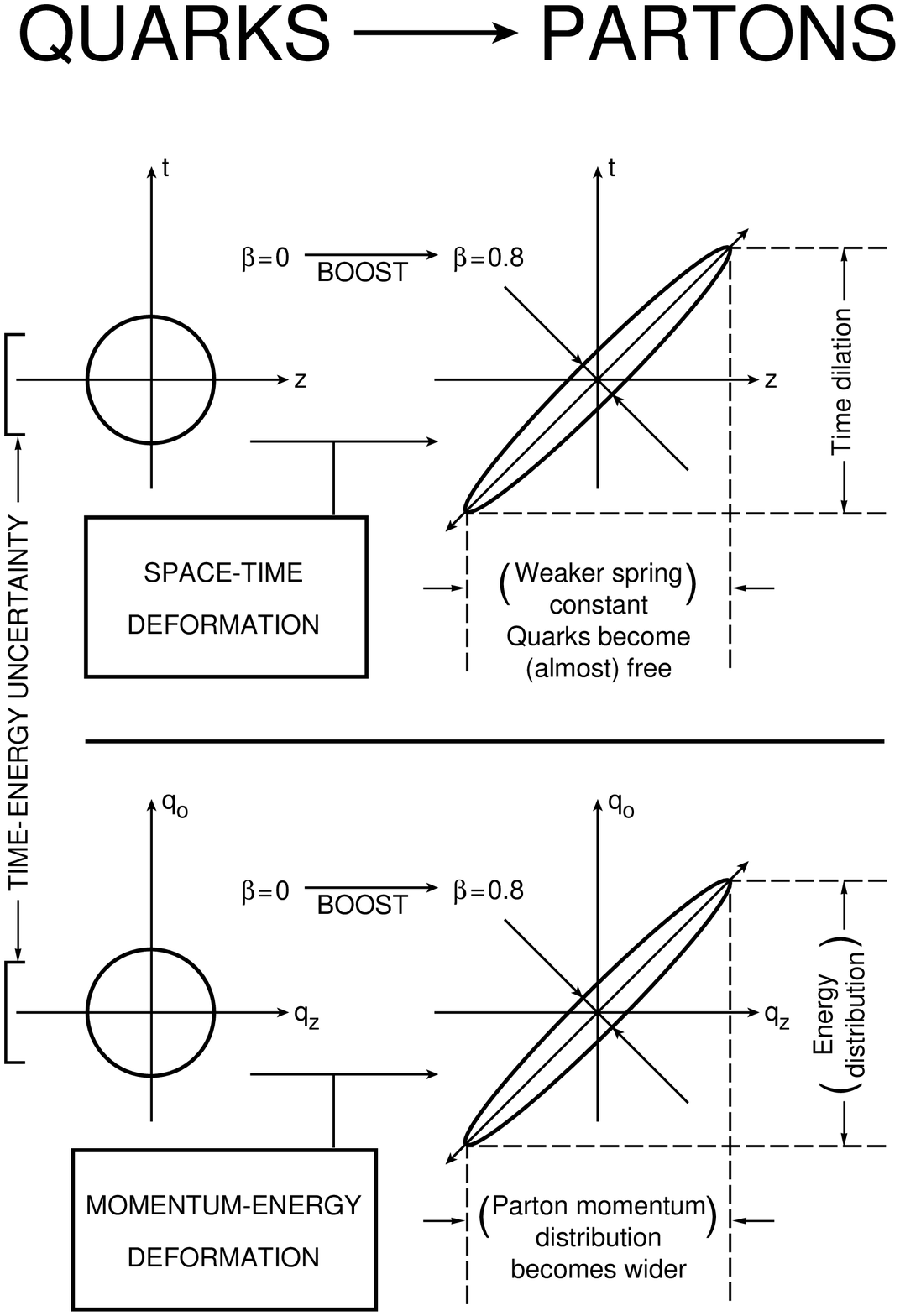}}
\vspace{5mm}
\caption{Lorentz-squeezed space-time and momentum-energy wave
functions.  As the hadron's speed approaches that of light, both
wave functions become concentrated along their respective positive
light-cone axes.  These light-cone concentrations lead to Feynman's
parton picture.}\label{parton}
\end{figure}

The momentum-energy wave function is just like the space-time wave
function, as is shown in Fig.~\ref{parton}.  The longitudinal momentum
distribution becomes wide-spread as the hadronic speed approaches the
velocity of light.  This is in contradiction with our expectation from
non-relativistic quantum mechanics that the width of the momentum
distribution is inversely proportional to that of the position wave
function.  Our expectation is that if the quarks are free, they must
have their sharply defined momenta, not a wide-spread distribution.

However, according to our Lorentz-squeezed space-time and
momentum-energy wave functions, the space-time width and the
momentum-energy width increase in the same direction as the hadron
is boosted.  This is of course an effect of Lorentz covariance.
This indeed is the key to the resolution of the quark-parton
paradox~\cite{knp86,kn77par}.

\begin{figure}
\centerline{\includegraphics[scale=0.6]{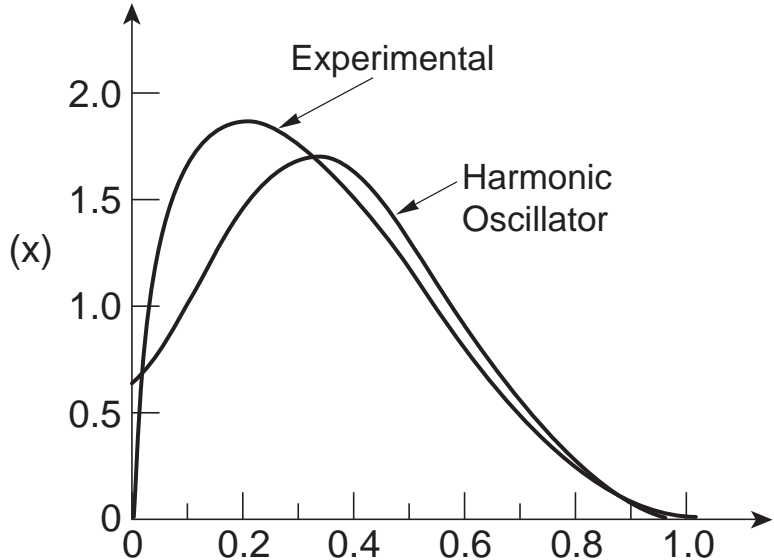}}
\vspace{5mm}
\caption{Parton distribution function.
Theory and experiment.}\label{hussar}
\end{figure}

After all these qualitative arguments, we are interested in whether
Lorentz-boosted bound-state wave functions in the hadronic rest
frame could lead to parton distribution functions.  If we start with
the ground-state Gaussian wave function for the three-quark wave
function for the proton, the parton distribution function appears
as Gaussian as is indicated in Fig.~\ref{hussar}.  This Gaussian  form
is compared with experimental distribution also in Fig.~\ref{hussar}.

For large $x$ region, the agreement is excellent, but the agreement is
not satisfactory for small values of $x$.  In this region, there is
a complication called the ``sea quarks.''  However, good sea-quark physics
starts from good valence-quark physics.  Figure~\ref{hussar} indicates
that the boosted ground-state wave function provides a good valence-quark
physics.

\section*{Concluding Remarks}

The present authors have been interested in question of covariant
harmonic oscillators since 1973~\cite{kn73}.  We started
with the covariant oscillator wave function as a purely
phenomenological mathematical instrument.  We then noticed that
the covariant oscillator formalism can serve as a representation
of the Wigner's little group for massive particles, capable of
the fundamental symmetry representation for relativistic particles.
This allows us to deal with the c-number time-energy uncertainty
relation without excitations.  Furthermore, the Lorentz-boosted
Gaussian wave function produces a parton distribution in satisfactory
agreement with experimental data.

What are then Feynman's contributions to this subject?  In addition
to the formulation of the parton picture, he suggested the use of
harmonic oscillator wave functions to understand bound-state problems
in the covariant regime.   Then where does the Feynman diagram stand
in his scheme?  Feynman diagrams start with plane waves which are
running waves.  Harmonic-oscillator wave functions are standing waves.
For standing waves, we have to take care of the covariance of boundary
conditions or spectral functions.  This is precisely what we are
reporting in this report.

It is gratifying to note that there is only one covariant quantum
mechanics for both scattering and bound states.  For scattering
states, we are dealing with asymptotically free waves, and Feynman
diagrams start with plane waves.  For bound states, we should start
with standing waves.  The harmonic oscillator wave functions constitute
the starting example.

It is our understanding that the purpose of string theory is to understand
the physics inside particles. Since particles are localized entities in
the space-time region, string theory is  necessarily a physics of standing
waves if we are to preserve the present form of quantum mechanics.  The
Lorentz covariance of the standing waves is the major issue in string or
brane theory.

\end{document}